\documentclass[prd,draft,nofootinbib,showpacs]{revtex4}

\usepackage{amsmath,amsfonts,amssymb}

\newcommand{\vev}{\textit{vev}}
\newcommand{\luv}{\Lambda_{\mathrm{UV}}}

\begin{document}

\title{The Higgs mechanism in Finsler spacetimes}
\author{Lorenzo Sindoni}
\email{sindoni@sissa.it}
\affiliation{SISSA/ISAS, via Beirut 2-4, 34014, Grignano, Trieste, ITALY\\
and INFN, sezione di Trieste}

\begin{abstract}Finsler geometry has been recently re-discovered as an interesting possibility to describe
spacetime geometry beyond Riemannian geometry. The most evident effect of this class of models is the prediction
of modified dispersion relations for particles moving in such backgrounds.
In this paper, we are going to consider the effects of modified dispersion relations on a gauge field theory
with spontaneous symmetry breaking (SSB) associated to a Higgs field. The percolation of higher dimensional, Lorentz violating operators to lower dimensional ones is discussed. We also discuss the issue of SSB in a mono-metric Finsler scenario like the one
associated to the so-called very special relativity.\end{abstract}

\pacs{04.60.Bc,11.30.Cp,11.30.Qc}

\maketitle

\section{Introduction}

The theory of General Relativity (GR) is based on the implementation of the Lorentz symmetry as a local symmetry (Local Lorentz Invariance, LLI). This amounts to the description of spacetime as a (pseudo-)Riemannian manifold, at the kinematical level,
while the classical dynamics is given by the Einstein's equations. 

Despite all the efforts, there is still no complete quantum theory of the gravitational field. A common feature of all the approaches is the indication that at scales comparable with the Planck length the description of spacetime as a (pseudo-)Riemannian manifold is at least inaccurate. Therefore, a new theory has to be used, whose nature we are not going to discuss here, which reproduces the geometrical picture of GR in a suitable regime. What can be said is that there will be at least a regime, between our low-energy scale and the Planck scale, in which we should still be
able to describe spacetime as a differentiable manifold, endowed with some new (effective) ``metric'' structure. Since LLI is deeply related to
the metric tensor, this means that abandoning pseudo-Riemannian geometry will bring us necessarily to a scenario in which LLI is
at least deformed, if not broken at all. In the past a lot of work has been done in this sense, both from the theoretical side
in the direction of Lorentz symmetry
deformation (deformed/doubly special relativity, DSR) \cite{Amelino-Camelia:2000mn, Magueijo:2001cr, Magueijo:2002am}, as well
as in the direction of building particle physics models with Lorentz Invariance Violation (LIV) effects included
\cite{Kostelecky:1988zi, Colladay:1996iz, Colladay:1998fq, Bluhm:2005uj, Bolokhov:2007yc}. These theoretical developments have been accompanied by the analysis of the constraints coming from experiments and astrophysical observations \cite{Coleman:1998ti, Mattingly:2005re, Jacobson:2005bg}, with increasing accuracy as long as new tests are proposed.

Here we are not dealing with a specific theoretical model which could give a solid ground on these speculations. Rather, we will assume that there is a modified geometrical description of spacetime, from which we are going to discuss the predictions for low-energy phenomenology through a suitably formulated effective field theory.
Since free particles have to move along some ``geodesics'' of the not yet specified geometry, to say that this is not (pseudo-) Riemannian geometry means that the considered ``geodesics'' (suitably determined in terms of the masses of the particles) are not the corresponding ones of (pseudo-)Riemannian geometry, or, equivalently, that the particles obey Modified Dispersion Relations (MDR).

The appearance of new terms in the dispersion relation has been suggested in different scenarios for quantum gravity \cite{AmelinoCamelia:1997gz, Gambini:1998it, Alfaro:1999wd, Dalvit:2000ay, Alfaro:2001rb, Burgess:2002tb, Alfaro:2002xz, Girelli:2006sc}. These MDR are produced through the introduction in the Lagrangian of new terms containing higher
order derivatives of the fields, which modify the equations of motion given by the Lorentz invariant Lagrangian. The important fact on which our discussion is based is that a MDR
can be seen as the manifestation of Finsler structure\footnote{See, for instance, \cite{rund,baochernshen} for introductions on Finsler geometry.} of spacetime \cite{Girelli:2006fw}, which therefore represents the geometrical theory corresponding to higher order derivatives theories. In this paper we are going to consider some implications
of a Finsler geometry structure of spacetime on particle physics phenomenology. It is important to remember that Finsler geometry has already been considered in the past in this kind of context, to cite only the works which have a direct connection with this discussion, as an alternative geometrical structure for Special Relativity in 1+1 dimensions \cite{Lalan, seb},
as well as anisotropic extensions to 3+1 dimensions \cite{Gonner:1997ec, Bogoslovsky:1999pp, Bogoslovsky:2007gt, Cohen:2006ky, Gibbons:2007iu},
as a possible way to avoid the GZK cutoff \cite{Kirzhnits:1972sg}, and as the most general spacetime geometry which can be realized
in an emergent geometry scenario \cite{Barcelo:2001cp, Visser:2001jd, Barcelo:2005fc, Weinfurtner:2006wt}.

Constraints on LIV in the Higgs sector have already been considered in \cite{Anderson:2004qi, Turan:2004iz}, at the level of the lowest dimensional operators of the standard model extension with LIV.
In this paper we want to take a different perspective, focusing on the way in which the Finsler structure can affect the low-energy phenomenology of a spontaneously broken gauge theory through an Higgs mechanism \cite{Englert:1964et, Higgs:1964pj, Guralnik:1964eu, Weinberg:1967tq}, including the effects of higher dimensional operators in a systematic
way. This problem is of a key relevance for the development of Finsler extensions of
the Standard Model (SM). The aim is to be able to propose (astro-)particle physics experiments to tests not just Lorentz invariance,
but spacetime geometry as Finsler geometry. Here we are doing only a first step in this program: in this work we are not going to
discuss the couplings with fermions, we limit the analysis to the gauge sector of the theory. 

In section II we are going to
consider the situation proposed by renormalization group (RG) arguments \cite{Girelli:2006sc}, in which we have that all the massless particles are described by the standard (pseudo-)Riemannian geometry Lagrangian, while the massive ones have a MDR.
If we assume that the Higgs has a MDR, we see that there is a non-trivial percolation of the LIV terms for the Higgs into lower dimensional operators
for the gauge bosons, which cannot be predicted from an analysis like the one proposed in the LIV standard model extension of \cite{Colladay:1998fq}, and whose suppression is ruled by the hierarchy between the UV energy scale responsible for
the Lorentz violating operators and the scale of symmetry breaking. In section III
we discuss the case of very special relativity (VSR) \cite{Cohen:2006ky, Gibbons:2007iu}, in particular pointing out some difficulties in 
realizing an Higgs mechanism in such Finsler scenarios. The outcome of the discussion is that particle physics 
can give some hints on which kind of geometrical structures we should prefer to use in model-building for phenomenological
scenarios for testing spacetime geometry beyond GR.

\section{Polynomial MDR}

In the effective field theory approach \cite{Myers:2003fd}, it is customary to introduce the MDR according to
the classification of the operators in terms of their canonical dimensions. This fits perfectly with MDR which admit
a polynomial expansion in terms of the momenta. This is possible, by dimensional arguments, if a dimensionfull quantity,
an energy scale, is introduced in the theory. Typically, this high energy scale related to LIV is identified with the 
Planck mass $M_{P}$, which is the scale at which new gravitational physics, and hence geometry, is expected to
relevantly modify our notion of spacetime. In what follows, instead of making such an assumption, we introduce a generic UV cutoff $\luv$ without specifying its origin. 
The main difficulty
in this approach is the large number of operators which must be added to the SM Lagrangian, at least in absence of some
guiding principle which can be used to restrict the possible additional terms to a specific class. 

In \cite{Girelli:2006sc} it has been shown that it is reasonable to expect that, due to RG effects, the geometry felt by
fields can be energy dependent. In particular, it was shown that, as a consequence, the various fields get a MDR according to
their mass: while massive fields have a modified mass shell, the massless ones do not, their mass shell being always the
light cone relative to the (low energy) Minkowski metric. 

These results can be used to introduce a specific class of models, with simple arguments based on symmetry principles. At high energy, the masslessness of gauge fields is protected by gauge invariance, which is unbroken in the high energy phase of the theory, and then, by the RG argument, their dispersion relation is unchanged. On the contrary, there is no symmetry protecting the Higgs Lagrangian from acquiring additional terms producing a MDR, since gauge invariance is not limiting enough the shape of the potential term, and in particular allows a mass term.

It is interesting, therefore, to discuss what happens in this scenario,
where the gauge fields have the standard kinetic terms, while the Higgs field's Lagrangian has a modification according to
the (yet to be) predicted MDR.

\subsection{Abelian Higgs model}
Let us consider the case of an abelian Higgs model. The Lagrangian for the Lorentz invariant case is given by\footnote{Conventions: the Minkowski metric is given by $\mathrm{diag}(+,-,-,-)$.}:
\begin{equation}\label{eq:lagabelhiggs}
    L= -\frac{1}{4}F_{\mu\nu}F^{\mu\nu} +
    \eta^{\mu\nu} D^{\dagger}_{\mu}\phi^{\dagger} D_{\nu}\phi -V(\phi),
\end{equation}
where $D_{\mu}=\partial_{\mu}+i g A_{\mu}$ is the covariant derivative containing the gauge field $A_{\mu}$, $F_{\mu\nu}=\partial_{\mu}A_{\nu}-\partial_{\nu}A_{\mu}$, $g$ is the gauge coupling and $V(\phi)$ is the potential,
which we consider of the form $-\mu^2 |\phi|^2 +\lambda|\phi|^4$ just to discuss a specific model, without loss of generality.

Let us add to scalar field Lagrangian a term which modifies the kinetic term, for example a $p^4-$like term
\begin{equation}
\frac{\eta}{\luv^2}  M^{\mu\nu\rho\sigma}(D_{\mu}D_{\nu}\phi)^\dagger D_{\rho}D_{\sigma}\phi,
\end{equation}
where $\eta \sim O(1)$, $\luv$ is the high energy scale corresponding to the physics generating this term and $M$ is
a tensor whose form is not specified here.
We have to consider what happens in the case when $\phi$ gets a vacuum expectation value (\vev). In particular, writing $\phi = ((v
/\sqrt{2})+\varphi)e^{i\theta}$, where $v=\mu/\lambda^{1/2}$ is the \vev, we get the mass term for the gauge field as usual, as well as an additional contribution coming from the MDR, which is easily obtained:
\begin{equation}
    D_{\mu}D_{\nu}\phi\rightarrow (\partial_{\mu}+ig A_{\mu})(\partial_{\nu}+i g A_{\nu})(v/\sqrt{2}+\varphi).
\end{equation}
This term generates new interactions between the (now massive) gauge boson $A_{\mu}$ and the field $\varphi$, which modify
the ones already present in the Lorentz invariant Lagrangian. Moreover, there is a whole
new part to be included in the action for the gauge boson alone, modifying its propagator. In particular, there is the term:
\begin{equation}
    \frac{\eta}{2} \left(\frac{gv}{\luv}\right)^2
M^{\mu\nu\rho\sigma}(\partial_{\mu}A_{\nu}-igA_{\mu}A_{\nu})(\partial_{\rho}A_{\sigma}+igA_{\rho}A_{\sigma}).
\end{equation}
This term goes directly into the renormalizable part of the action related to the gauge boson.
Notice that this amounts to a new quartic self-interaction governed by the tensor $M^{\mu\nu\rho\sigma}$, which includes the
effect of Lorentz violation, and, most important, the modification of the kinetic term at the level of dimension four
operators. We recognize the combination $g v=M_A$ is nothing but the mass of the gauge field.

For different MDR for the Higgs field, the discussion is similar. For example, for a $p^{2n}$ modification\footnote{The discussion of $p^{2n+1}$ modifications, is exactly the same, even though it is clear that for that class one should consider necessarily the fate of the invariance of the Lagrangian under discrete symmetries, $C,P,T$. For the simplicity of the discussion, here we consider only even powers of the
momenta.}, we get a
$p^{2n-2}$ modification of the kinetic term for the massive gauge boson, inherited from the scalar field, as well as
new self-interactions.

These terms are highly constrained from astrophysical observations for particles like photons and electrons ({\it i.e.} the QED sector), but not for the bosons $W^{\pm},Z^0$, for which the analysis of Lorentz invariance have not been considered yet, at least with the same accuracy. As it stands, however, this discussion is not completely satisfactory for a phenomenological analysis, since we are still discussing the abelian case, while we should discuss the most general case of nonabelian gauge theories. What we can conclude, at this stage, is that LIV in the form of a MDR is propagating in
the Lagrangian of a spontaneously broken gauge theory in a non trivial way even at the tree level, without taking into
account quantum corrections, whose role could be even more important \cite{Collins:2004bp}.

However, despite being potentially relevant, besides being suppressed by suitable powers of the cutoff $\luv$ required by dimensional analysis,
all the new terms are multiplied by the dimensionless ratio $r=M_A ^2/\luv^2$, between the square of the mass of the gauge boson and the square of the UV cutoff. If the MDR were related to the Planck
length and $v \approx \mathrm{TeV}$ the electro-weak (EW) scale, we could conclude that $r \approx 10^{-32}$, thus enhancing the
Planck suppression through the large hierarchy between the EW scale and the Planck scale.

The bottom line of this discussion is that the modification of the dispersion relation induced on the gauge boson is more suppressed than expected, and thus we can detect it only in extremely accurate precision tests of our models\footnote{Therefore,
to accomplish this task one should move beyond the tree level analysis and consider the first quantum corrections to the SM Lagrangian.}.

\subsection{Non-Abelian}

To fully understand the implications of a MDR in a realistic model for particle physics, we have to discuss the case of non-abelian
gauge fields. Let us consider the case of a non-abelian gauge group, like for instance an $SU(N)$ gauge theory, with generators of the Lie algebra given by the matrices $T_{A}$, and gauge fields $G_{\mu}^{A}$.
Let us consider a Lorentz invariant Higgs model:

\begin{equation}\label{eq:lagnonabelhiggs}
    L = -\frac{1}{4} \mathrm{Tr}(G_{\mu\nu} G^{\mu\nu}) + |D_{\mu}\Phi|^2 - V(\Phi).
\end{equation}

As before, let us add a MDR term, $p^4-$like, and discuss what happens if the Higgs multiplet gets a \vev. The term
to be added can be written as:

\begin{equation}
\frac{\eta}{\luv^2}M^{\mu\nu\rho\sigma}(D_{\mu}D_{\nu}\Phi)^{\dagger}D_{\rho}D_{\sigma}\Phi.
\end{equation}

If the potential $V$ allows a non-vanishing \vev~of the multiplet $\Phi$, so to break spontaneously the gauge symmetry,
we have the gauge bosons mass terms generated by the standard kinetic term of the Higgs
and, as before, new contributions coming from the additional term encoding the MDR. Now:
\begin{widetext}
\begin{equation}
    D_{\rho}D_{\sigma}\Phi\rightarrow (\partial_{\rho} + i g T_A W^{A}_{\sigma})(\partial_{\sigma} +i g T_A W^{A}_{\rho})(\langle\Phi\rangle+\varphi).
\end{equation}
Neglecting the terms describing the interaction of the gauge bosons with the field $\varphi$, we have a new part
for the action of the gauge bosons alone:
\begin{equation}
    M^{\mu\nu\rho\sigma}[(-i T_{A}\partial_{\mu}W^A_{\nu}-g T_AT_BW^{A}_{\mu}W^{B}_{\nu})V]^{\dagger}[(+i T_{C}\partial_{\rho}W^C_{\sigma}-g T_CT_DW^{C}_{\rho}W^{D}_{\sigma}V)],
\end{equation}
\end{widetext}

These new terms include new self-couplings to be added to those given by the original Lagrangian, plus a dimension four
operator which has to be added to the standard kinetic term and which modifies the dispersion relation of the gauge
bosons at the quadratic level, as in the abelian case.

The power counting argument on the strength of the modification is left unchanged. The structure is otherwise
the same as in the abelian case. In principle, we could expect three features:
\begin{itemize}
  \item (additional) mixing between different gauge bosons induced by the MDR;
  \item modification of the dispersion relation at the level of $p^{2n-2}$ instead of $p^{2n}$;
  \item additional three and four gauge bosons interactions.
\end{itemize}
All these points must be taken into account as potential sources for new physics, beyond the predictions of SM.

Without loss of generality, let us consider the special case of a spontaneously broken $SU(2)\times U(1)$ gauge theory, with an Higgs doublet, which
is relevant for the SM dynamics. To avoid confusion on conventions, we write down step by step the Lagrangian, in order
to make the comparison with the standard case easier.
The part of the Lagrangian involving only the Higgs doublet and the gauge fields is given by:
\begin{widetext}
\begin{equation}\label{eq:gswlag}
 L_{0} = -\frac{1}{4}\mathrm{Tr}(\mathbf{F}_{\mu\nu}\cdot \mathbf{F}^{\mu\nu}) - \frac{1}{4}G_{\mu\nu}G^{\mu\nu} +
(D_{\mu}\Phi)^{\dagger}D^{\mu}\Phi+ \mu^2 \Phi^\dagger \Phi - \frac{\lambda}{4}(\Phi^{\dagger}\Phi),
\end{equation}

where the SU(2) gauge fields, $W^{i} _{\mu}$, and the U(1) gauge vector $B_{\mu}$ have field strengths given
respectively by:

\begin{equation}
 \mathbf{F}_{\mu\nu} = \partial_{\mu}\mathbf{W}_{\nu}-\partial_{\nu}\mathbf{W}_{\mu}-g \mathbf{W}^{\mu}\times\mathbf{W}^{\nu}, \qquad
G_{\mu\nu}=\partial_{\mu}B_{\nu}-\partial_{\nu}B_{\mu},
\end{equation}
\end{widetext}
where we have used the compact notation $\mathbf{W}^{\mu}= \mathbf{\tau}_i W^i_{\mu}$, with $\tau$ the Pauli matrices, and where
$g,g'$ are the two dimensionless coupling constants. The covariant
derivative to be applied on the Higgs doublet is given, as usual, by:
\begin{equation}
 D_{\mu} = \partial_{\mu} + i\frac{g}{2} \mathbf{W}_{\mu} + i \frac{g'}{2} B_{\mu}.
\end{equation}
Working in the unitary gauge, we parametrize the \vev~of the doublet in the form:
\begin{equation}
 \langle \Phi \rangle = \left( \begin{array}{c}
 0 \\
v/\sqrt{2}
\end{array}
 \right),
\end{equation}
whence the standard massive gauge bosons Lagrangian can be easily computed. Let us now suppose that the Higgs doublet, due to
fuzziness of spacetime at small scales, shows a modified dispersion relation of the $p^4$ kind and therefore let us add to
\eqref{eq:gswlag} the term
\begin{equation}\label{eq:Higgsp4}
 L_{p^4}=- \frac{\eta}{\luv^2} M^{\mu\nu\rho\sigma}\left(D_{\mu}D_{\nu}\Phi^{\dagger}\right)D_{\rho}D_{\sigma}\Phi.
\end{equation}

Clearly, the addition of this term has no effect on the value of the \vev, but it does have an effect on the shape
of the kinetic terms for the gauge bosons, as well as new interactions. Let us neglect this last issue, and let us focus on the
contributions to the kinetic terms. The additional terms are easily identified since they come from the terms
\begin{equation}
 \partial_{\mu}(D_{\nu}\langle\Phi\rangle).
\end{equation}
After some algebra, we conclude that, in addition to the standard kinetic terms and to the dynamically generated mass terms, the free part of the Lagrangian for the gauge bosons contains
the following contribution:

\begin{equation}
 - \eta g^2 \frac{ v^2}{\luv^2} M^{\mu\nu\rho\sigma}\left[ (\partial_{\mu} W_{1\nu} +i\partial_{\mu}W_{2\nu})
(\partial_{\rho} W_{1\sigma} -i\partial_{\rho}W_{2\sigma}) +
\left(\frac{g'}{g} \partial_{\mu} B_{\nu} -\partial_{\mu}W_{3\nu}\right)
\left(\frac{g'}{g}\partial_{\rho} B_{\sigma} - \partial_{\rho}W_{3\sigma}\right) \right].
\end{equation}
As it is easily seen, the mixing terms between the fields $B$ and $W_3$ can be removed in the standard way, if we introduce the combinations $Z^{\mu},A^{\mu}$
\begin{equation}
  Z^{\mu} = \cos{\theta_{W}} W_3^{\mu} - \sin{\theta_{W}}B_{\mu},
\end{equation}
\begin{equation}
 A^{\mu} = \cos{\theta_{W}} B^\mu+\sin{\theta_{W}} W_3^{\mu},
\end{equation}
with
\begin{equation}
 \cos{\theta_{W}} = \frac{g}{(g^2+g'^2)^{1/2}}, \;\;\;\;  \sin{\theta_{W}} = \frac{g'}{(g^2+g'^2)^{1/2}},
\end{equation}
and therefore the additional term\footnote{Here we assume that the matrix $M$ is real.} coming from \eqref{eq:Higgsp4} after symmetry breaking becomes
\begin{equation}\label{eq:bosonmdr}
 -\eta g^2 \frac{v^2}{\luv^2} M^{\mu\nu\rho\sigma}\partial_{\mu}W_{1\nu}\partial_{\rho}W_{1\sigma}
 -\eta g^2 \frac{v^2}{\luv^2} M^{\mu\nu\rho\sigma}\partial_{\mu}W_{2\nu}\partial_{\rho}W_{2\sigma}
 -\eta (g^2+g'^2) \frac{v^2}{\luv^2} M^{\mu\nu\rho\sigma}\partial_{\mu}Z_{\nu}\partial_{\rho}Z_{\sigma}.
\end{equation}
The particle spectrum is easily deduced. The masses are the same of the Lorentz-invariant case: there are two massive gauge bosons, $W_{1,2}$, which have mass given by $M_W= gv/2$, a $Z$ boson with $M_Z = M_W/\cos\theta_W$ and a massless gauge field which
represents the electromagnetic field, associated to the residual $U(1)$ gauge invariance of the model. However, while
this residual gauge invariance protects the $A_{\mu}$ from dangerous terms containing the (Lorentz violating) tensor $M^{\mu\nu\rho\sigma}$, the other gauge bosons have a modified dispersion relation at the level of dimension four operators
given by
\begin{equation}\label{eq:wmdr}
 -4 \eta \frac{M_W ^2}{\luv^2} M^{\mu\nu\rho\sigma}\partial_{\mu}W_{\pm \nu}\partial_{\rho}W_{\pm \sigma},
\end{equation}
for the bosons $W^{\pm}$, while for the $Z^0$ the modification is given by:
\begin{equation}\label{eq:zmdr}
 -4 \eta \frac{M_Z ^2}{\luv^2} M^{\mu\nu\rho\sigma}\partial_{\mu}Z_{ \nu}\partial_{\rho}Z_{ \sigma}.
\end{equation}
Notice that this modification is mass dependent: the $W^{\pm}$ bosons will receive a modification which is smaller of the one for the $Z^0$.

Higher order derivative operators contribute in a similar way. Let us consider, for instance, a term like:
\begin{equation}
 \frac{\eta_{(n)}}{\luv^{2n-2}}M^{\mu_1...\mu_{2n}}(D_{\mu_1}... D_{\mu_{n}}\Phi)^{\dagger}D_{\mu_{n+1}}...D_{\mu_{2n}}\Phi.
\end{equation}
When the Higgs gets a \vev, the modification of the kinetic term for the gauge bosons is easily seen to be
\begin{equation}
 \frac{\eta_{(n)}}{\luv^{2n-2}}M^{\mu_1...\mu_{2n}}(\partial_{\mu_1}... \partial_{\mu_{n-1}} D_{\mu_{n}}\langle\Phi\rangle)^{\dagger}\partial_{\mu_{n+1}}...\partial_{\mu_{2n-1}}D_{\mu_{2n}}\langle\Phi\rangle.
\end{equation}

In particular, the matrix structure is the same as the $p^4$ modification, and therefore we can conclude that the photon Lagrangian will not get modifications, while the massive gauge bosons will receive $p^{2n-2}$ modifications to their propagators, which will be
suppressed by the ratio $(M^{2}_{\mathrm{boson}}/\luv^2)$, besides the standard suppression given by powers of the UV cutoff. 
In general, the MDR will not change the diagonalization procedure necessary to extract the mass eigenstates representing the physical propagating modes: they will be given by the same combinations as in the Lorentz invariant case. What is different is just the
shape of the dispersion relation/kinetic term. In the specific case we have considered, the $SU(2)\times U(1)$,
there was no modification at all of the Weinberg's angle. This is totally general, being related to the
fact that the kinetic term is a field bilinear: adding derivatives we do not touch the matrix structure, hence
we do not introduce extra sources of mixing, as we naively expected. 

The final outcome of this discussion is pretty easy to understand: the
MDR of the Higgs propagates in the Lagrangian of the massive gauge bosons in such a way to produce a MDR which is not of the
type which we are inserting at the beginning. In particular, the corrections in the form of $p^{2n}$ operators for the Higgs become
effectively $p^{2n-2}$ terms for the massive gauge bosons. Nevertheless, as we have shown, these
modifications to standard model are further suppressed by $M_{\mathrm{boson}} ^2/\luv^2$, which means that if the SSB scale and the
LIV scale are too far away these terms are negligible, at least at the classical level. This
large suppression can make these new terms still compatible with present constraints on dimension four operators: larger modifications would be already ruled out. 

The modification to the MDR of the massive gauge bosons Lagrangian is polarization dependent. This can be understood easily since gauge symmetry is broken, and since the would-be Goldstone bosons coming from the Higgs multiplet are included in the gauge fields corresponding to the broken generators, becoming their
longitudinal component, have a different dispersion relation with respect to
the transverse polarizations. This ultimately results in a polarization dependent dispersion relation.
Correspondingly, the residual gauge invariance $U(1)_{em}$ protects the photon
from acquiring Lorentz violating terms, at least at the tree level. 

Therefore, in order to present an extension of the SM taking into
account a sort of energy-dependence geometrical structure of spacetime, when an Higgs mechanism is invoked to have SSB, the
analysis of the Lagrangian must be done with care, since new terms appear which cannot be expected naively
from just the basic principles one is using, like gauge invariance. Moreover, the extra suppression given by the dimensionless ratio $(M_{\mathrm{boson}}/\luv)^2$ cannot be obtained from dimensional analysis alone.


\section{Very Special Relativity}

The discussion of the Higgs model in the RG setup has highlighted that a MDR for the Higgs field does not produce a MDR of the same kind for the gauge boson. This means that to really believe that a MDR is the manifestation of a modified geometrical structure which wants to be universal as (pseudo-)Riemannian geometry is for SM, then it should at least
be compatible with an Higgs mechanism. If not, in order to save the fundamental role of geometry, we must find an
alternative scheme of SSB which is compatible with the particular geometrical structure.

In the previous section we have considered a situation in which the geometrical structure is particle dependent.
It is interesting, therefore, to consider now a specific model of a ``universal'', particle independent, modified geometrical structure and to consider on this
background the simplest version of a spontaneously broken gauge theory.

Very special relativity has been proposed as a theory in which relativistic invariance is reduced due to the presence of a preferred null vector field.
In VSR, the Finsler line element is\footnote{For the details for the formulation of a field theory in this case, we refer to \cite{Gibbons:2007iu}.}
\begin{equation}\label{eq:VSRfinsler}
    ds^2= (\eta_{\mu\nu}\dot{x}^{\mu}\dot{x}^{\nu})^{1-b}(n_{\mu}\dot{x}^{\mu})^{2b},
\end{equation}
where $n_{\mu}$ is a constant null vector field, and $b$ a real parameter. The corresponding modified dispersion
relation is:
\begin{equation}\label{eq:VSRMDR}
    (\eta^{\mu\nu}p_{\mu}p_{\nu})^{1-b}(n^{\mu}p_{\mu})^{2b} = m^2,
\end{equation}
where we are raising and lowering indices through the Minkowski metric $\eta_{\mu\nu}$.

The symmetry group which leaves invariant this Finsler line element is a subgroup of the Weyl group which leaves invariant the direction of the four vector $n^\mu$, besides leaving invariant the Minkowski metric, up to rescalings. This group has eight generators: four translations, a combination of the boost along $n$ and the identity $N_{n}-\xi \mathbb{I}$, the rotation around the spatial part of $n$ $J_{n}$, and two combinations of the boosts and rotations in the transverse directions\footnote{See for instance \cite{Gibbons:2007iu} for an accurate discussion of the details. For a slightly different perspective we refer to the works of G. Y. Bogoslovky, see, for instance \cite{Gonner:1997ec,Bogoslovsky:1999pp}}. 
 
For massless particles, the dispersion relation is just the special relativistic one. Notice that, despite
its Finsler nature, the line element \eqref{eq:VSRfinsler} is just obtained from the Minkowski one with a (Finsler-like) conformal transformation. In particular, the causality relations are the same as in special relativity.

Let us suppose that we have an abelian
Higgs model on this Finsler spacetime. The Lagrangian will be:

\begin{widetext}
\begin{equation}\label{eq:VSRhiggslag}
    L=-\frac{1}{4}F^{\mu\nu}F_{\mu\nu} - \phi^{\dagger} (\eta^{\mu\nu}D_{\mu}D_{\nu})\phi + m^2\phi^{\dagger}\left(\frac{in^{\mu}D_{\mu}}{m}\right)^{B} \phi  -\frac{\lambda}{2} |\phi|^4,
\end{equation}
\end{widetext}
where the kinetic term of the scalar field is determined by the MDR through the replacement $p_{\mu}\rightarrow i\partial_{\mu}$ and then asking that gauge invariance holds, replacing partial
derivatives with gauge covariant derivatives. Here we have introduced the notation $B=2b/(1+b)$ to make expressions simpler to manipulate. Notice that this Lagrangian is the only one compatible with gauge invariance and the VSR relativity group \cite{Gibbons:2007iu}.

Clearly this Lagrangian has a non-polynomial form which could make
difficult an explicit treatment of the interactions, even at the perturbative level.
By dimensional analysis,
since there is no energy parameter entering in the dispersion relation which could be used as an
expansion parameter to produce a series of differential operators of increasing order, the truncation
this kinetic term at a given order of differentiation is impossible.

The outcome of the SSB is the appearance of massive gauge bosons. This implies that we
have to obtain a term which reads:
\begin{equation}\label{eq:massterm}
    m^2\int d^4k \,\eta^{\mu\nu}\tilde{A}_{\mu}(-k)\left( \frac{n^{\rho}k_{\rho}}{m} \right)^{B} \tilde{A}_{\nu}(k),
\end{equation}
in momentum space. This term has the correct transformation properties with respect to the spacetime symmetry group
to represent the mass term appropriate for the given dispersion relation.

To see what happens in this Finsler setting, let us consider in detail the case of the free scalar field with the only
potential which is allowed by the relativistic symmetry group of the model.
The field equation, neglecting the possible coupling with the gauge field, is:
\begin{equation}\label{eq:scalfieldeq}
    \Box \phi -\mu^2 \left( i\frac{n^{\rho}\partial_{\rho}}{\mu} \right)^{B} \phi + \lambda |\phi|^2 \phi=0,
\end{equation}

Notice that this model does not have a smooth limit when $b\rightarrow 0$, since the function $x^a$ is not analytic
in $x=0$, if $a$ is not a positive integer. In the case $b=0$, which corresponds to
the special relativistic case,
this equation admits the constant solutions:
\begin{equation}
    \phi = v e^{i\theta },
\end{equation}
with $v= \mu/(2\lambda)^{1/2}, \theta \in [0,2\pi)$. However, if $b\neq 0$, it is easy to see that the only constant solution has $v=0$. This is consistent with the fact that one of the boosts is mixed with a dilatation, which does not leave the field
$\phi$ invariant. Therefore, the only constant solution which is compatible with the transformation properties of the
field under the relativity group is the identically vanishing solution. A \vev~ for $\phi$ would break the VSR
group to a smaller group, in the specific case, the subgroup of $SO(3,1)$
which leaves invariant the vector $n^{\mu}$. Despite being a logical possibility, the link with the Finsler line element
\eqref{eq:VSRfinsler} would be weakened, since it is true that it is left invariant by this smaller group,
but the Minkowski line element would be left invariant as well.

We can conclude then that an Higgs mechanism with an Higgs field taking a constant \vev~is incompatible with the VSR scenario,
whose spacetime symmetry group forbids in the Lagrangian any operator which would be able to produce such a constant expectation value. It is worth noting that even a scenario in which a fermion-antifermion condensate is formed, $\langle
\bar{\psi}\psi\rangle$, is problematic for the same reason. Under the relativity group of VSR, the wave function of the
fermion undergoes dilatations, again making the condensation mechanism incompatible with the relativity group. This means that the problems we are encountering are quite independent from the specific model adopted.

The only other alternative to this scenario is the case of the condensation of some other operator, which has the dimensions of a squared mass, and which is allowed by the symmetries of the system, in particular a scalar which can
define uniquely the mass of the gauge boson in every reference frame connected by a relativity transformation.
It is easily seen that an operator which could do the job is given by:

\begin{equation}\label{eq:operator}
    \mathcal{O}=\phi^{\dagger}\left(\frac{in^{\alpha}\partial_{\alpha}}{\mu}\right)^{-B}\phi.
\end{equation}
In particular, it is easy to see that the equation of motion \eqref{eq:scalfieldeq} has the following plane wave solutions:

\begin{eqnarray}
     \phi = v(k)e^{-ik_{\alpha}x^{\alpha}}, & \nonumber \\v^2(k) = \frac{1}{\lambda} \left[ \eta^{\alpha\beta} k_{\alpha}k_{\beta} - \mu^2 \left( \frac{n^{\alpha}k_{\alpha}}{\mu}\right)^{B} \right]. \label{eq:planewave}
\end{eqnarray}
Notice that this $v(k)$ has the correct transformation properties under the relativity group to represent
the amplitude of a scalar field. In the special case in which $k^2=0$, after elementary algebra one sees that
\begin{equation}
    \mathcal{O} = - \frac{\mu^2}{\lambda},
\end{equation}
which is left invariant by all the reference frame transformations considered.
This kind of operators can be therefore used to build interaction terms with the gauge fields to produce, in
certain regimes of the theory, a mass term for the gauge bosons. However, to do so, we have to deeply modify the
Lagrangian for the would be gauge field in a way which includes couplings which cannot be obtained just with
the minimal coupling prescription $\partial_{\alpha}\rightarrow D_{\alpha}$.

A potential difficulty is that in order for the Lagrangian to be gauge invariant,
the field $A_{\alpha}$ must enter either through the field strength $F_{\alpha\beta}$, or through the covariant
derivative $D_{\alpha}$. In particular, in order to get a mass term for a linear equation of motion of the massive gauge boson,
we need an operator containing two covariant derivatives, at most. To saturate the vector indices, we need a metric
tensor. Finally, for the Lagrangian to be a $U(1)$ scalar we need the combination $\phi^{\dagger} \phi$. The term just
described is nothing but:
\begin{equation}
    (iD_{\alpha} \phi)^{\dagger}(iD_{\beta}\phi)\eta^{\alpha\beta},
\end{equation}
which is already in the Lagrangian and cannot produce the desired mass term. Any other kind of operator can only
involve fractional derivatives, and therefore cannot give rise to bilinear expressions in the field $A_{\alpha}$,
but to highly non-polynomial operators like, for example,

\begin{equation}
\eta^{\alpha\beta}(iD_{\alpha}\phi)^{\dagger}\left(\frac{in^{\gamma}D_{\gamma}}{\mu}\right)^{-B}(iD_{\beta}\phi),	
\end{equation}
whose physical content, in terms of Feynman diagrams, is not clear at all.

A crucial difficulty, which makes this approach useless, is the fact that a background solution with a generic $k_\mu\neq n_{\mu}$ automatically breaks the VSR group, since the only vector which is left covariant by the symmetry group
is $n_{\mu}$ itself. It is interesting to note that, if we consider this latter
possibility, since $n_{\mu}$ is a null vector, we obtain that the corresponding amplitude $v(n)$ given by formula \eqref{eq:planewave} actually vanishes, therefore making impossible for our program to have a successful conclusion of generating massive gauge bosons via gauge symmetry rearrangement.

\section{Discussion}

We have described the interplay of the Higgs mechanism with MDR. We have shown that a MDR for the Higgs field corresponds
to new physics in the gauge sector of the theory, which could be tested, in principle, by precision tests in the massive gauge bosons sector. We have seen that, however,
in the most simple cases of polynomial dispersion relations the new physics is additionally suppressed by the hierarchy
between the SSB scale and the high energy scale which is associated to the dispersion relation. On one side, this makes them quite difficult to detect, but on the other side, this means that they could be still compatible with present
bounds on Lorentz symmetry violation. In particular this is relevant for dimension three and four operators, for which the
ratio $r$ is crucial to make them small enough to be compatible with observations. 

In this respect, and we stress again this point, the effect of SSB
is to translate an order one LIV effect in the Higgs sector into a largely suppressed LIV effect in the massive gauge bosons sector, this suppression being naturally small since it is just the square of the ratio between the EW scale and the Planck scale. Even though radiative corrections could modify this analysis, we can say that the Higgs mechanism can naturally realize a scenario in which LIV is very suppressed.\footnote{Notice that we have
just turned the problem of the smallness of the coefficients to the hierarchy problem: $M_{EW}/M_P \approx 10^{-16}$, which
requires an independent explanation.}.

Besides the large suppression, it is worth to note that, for instance, the
dimension four operators obtained as described in section II from dimension six operators after symmetry breaking, are not gauge
invariant: therefore, they cannot be naively predicted from the standard arguments \cite{Colladay:1998fq, Myers:2003fd}, for
which one uses operators which are gauge invariant. Of course, being not gauge invariant, these operators can appear only
for the gauge bosons corresponding to the broken symmetries, leaving untouched the photon's and the gluons' Lagrangian. For these fields, the source for Lorentz violating terms (if there is any LIV for these fields) must be independent: for instance, we can consider
models in which also the massless fields do have a MDR.

For what concerns phenomenological investigations,
these effects could be negligible for the EW theory: the EW scale and the high energy scale associated with the LIV, the Planck scale for instance, are widely separated, and therefore the massless ration $(M_{EW}/\luv)^2$ is extremely tiny.
In a GUT scenario, where the difference between the GUT scale
and the high energy scale $\luv$ can be significantly smaller, the effects could be more evident,
at least in principle. However, an experimental detection requires the analysis of the physics of the massive bosons coming from the breaking of the GUT symmetry, which is not accessible for our present technology.

Despite producing a potentially interesting phenomenology,
this class of MDR has the important theoretical drawback
of being described by a family of Finsler geometries, parametrized by the mass of the particles one is considering \cite{Girelli:2006fw},
losing the uniqueness of the geometrical background: it is impossible
to find a unique Finsler metric describing these new kinetic terms. Consequently, it is difficult to look at these situations as coming from a coherent underlying theory which should describe a ``semiclassical'' structure of spacetime.

It is interesting, then, to see what happens in a fully consistent Finsler setting, in which all the particles see a single
Finsler metric. In the specific case of VSR, we have shown that the SSB mechanism \textit{\`a la} Higgs cannot
be realized in the usual way, without breaking (very special) relativistic invariance. This is interesting
because it is a spontaneous symmetry breaking of a spacetime symmetry through a scalar \vev.  Moreover, the transformation law for the scalar field under certain changes of reference frame involves a dilatation factor, which is a sort of global $U(1)$ transformation with complex parameter, which is a subgroup of the gauge group of the theory. Touching
the gauge symmetry necessarily touches the relativistic symmetry: they are deeply entangled, in this specific model.

Even without doing explicit calculations to check the radiative corrections, it is interesting to note that a VSR field theory, containing only spinors interacting through gauge fields, at the massless level, has an enhanced degree of symmetry: it is a conformal theory, invariant under the whole Weyl group, since the vector $n^{\mu}$ never appears in the Lagrangian\footnote{In \cite{Bogoslovsky:2004rp, Gibbons:2007iu} the authors considered nonlinear self-interactions like $(\bar{\psi}\gamma^{\alpha}n_{\alpha}\psi)^b (\bar{\psi} \psi )^{1-b}$, but this is a non-polynomial term whose physical
meaning is not clear. Certainly, it does not describe the mass term of free spinors, since it leads to an equation of motion which is nonpolynomial.}. Nevertheless, we already know that scale invariance is broken by quantum corrections. In the
case of very special relativity, this would amount to the breaking of the relativistic symmetry to a subgroup of the
Lorentz group, as we have already discussed. This, on the other side, would spoil the Finsler line element \eqref{eq:VSRfinsler} of its privileged role as
the only line element left invariant by the relativity transformations. For instance, since the symmetry would be reduced to a subgroup of the
standard Lorentz group, the standard Minkowski line element is left invariant too, and one could introduce the preferred vector $n^{\mu}$
in other ways, which are not directly linked to a Finsler norm\footnote{Notice, however, that strictly speaking Riemannian
geometry is a special case of Finsler geometry.}.

In general, the discussion of radiative corrections can be crucial, in LIV scenarios \cite{Collins:2004bp}, since very suppressed higher dimensional operators can percolate on dimension three and four operators, when taking into account higher loops corrections. Here the situation is the same. Apart the SM vertices, which can be consistently renormalized, the new gauge bosons interactions can be particularly dangerous, since loops can amplify them. For example, in the simple scenario we have discussed in section II where a $p^4$ modification was considered, the resulting four bosons interaction will have a dimensionless coupling given by
$M_{EW}^2/\luv^2$. The relevance of this coupling changes dramatically when we include this vertex in the calculation of the self energy of a gauge boson: by simple arguments we see that the contribution becomes of the order of the EW scale:
\begin{equation}
  \frac{M_{EW}^2}{\luv^2} \int^{\luv} \frac{d^4k}{k^2-M_{EW}^2} \simeq \frac{M_{EW}^2}{\luv^2}\luv^2 = M_{EW}^2,
\end{equation}
without any further suppression.
In order to protect lowest order operators from these dangerous radiative corrections, we need some form of custodial symmetry which compensates this kind
of contribution with another one, with the opposite sign, for example. This can
be implemented, for example, providing a SUSY extension of the theory \cite{GrootNibbelink:2004za, MattinglyQGPH07}: a fermionic loop with the same amplitude but opposite sign would cancel this dangerous ``order one'' radiative correction.

At this point a comment on fermions must be made. If we suppose that the mass generation mechanism for them is given through Yukawa couplings with the Higgs, we see that there is no (tree-level) LIV/MDR induced by the Higgs. Of course, it
is conceivable that fermionic fields acquire directly a MDR due to QG effects, without necessarily passing through the Higgs, and of course, loop corrections will produce as well modifications to the propagators. 

To conclude, it is clear that the Higgs mechanism fits particularly well in Lorentz invariant theories, while it is difficult
to reconcile with different geometrical structures preserving their fundamental role. In particular, it is well designed to generate masses for gauge bosons in Lorentz-invariant gauge theories, while destroys the
geometrical interpretation in a Finsler background one like in VSR. The key point is that, while in the SM all the kinetic
terms are formed using field bilinears and at most two (gauge covariant) derivatives, MDR require more complicated expressions which are not trivial to manipulate when we introduce the
decomposition of the Higgs field into the \vev~and fluctuations. Conversely, if we want to preserve a unique
geometrical background, we need another mechanism for SSB which avoids this difficulty. 

Looking at the problem from a different and more ambitious perspective we could say that
an accurate study of the properties of the electroweak symmetry breaking could shed light onto spacetime structure beyond GR,
even though only in a very indirect way.

\acknowledgments{The author would like to thank Florian Girelli and Stefano Liberati for useful discussions and
constructive criticisms on earlier versions of the draft.}

\end{document}